\documentclass[preprintnumbers,amsmath,amssymb,floatfix,10pt,prd,superscriptaddress,nofootinbib,twocolumn]{revtex4}

\usepackage{doi}%<----------
\hypersetup{
  colorlinks=true,        % false: boxed links; true: colored links
  linkcolor=magenta,         % color of internal links
  citecolor=red,         % color of links to bibliography
}

\usepackage{graphicx}% Include figure files
\usepackage{dcolumn}% Align table columns on decimal point
\usepackage{bm}% bold math
\usepackage{color}
\usepackage{enumitem}
\usepackage{amsmath}
\usepackage{amssymb}

\newcommand{\beq}{\begin{equation}}
\newcommand{\eeq}{\end{equation}}
\newcommand{\bea}{\begin{eqnarray}}
\newcommand{\eea}{\end{eqnarray}}

\usepackage{bbm}
\usepackage{amsfonts}
\usepackage{mathrsfs}
\usepackage{latexsym}
\usepackage{epsfig}
\usepackage{epstopdf}
\usepackage{epstopdf}
\usepackage{graphicx}
\usepackage{amssymb}
\usepackage{amsmath}
\usepackage{dcolumn}
\usepackage{bm}
\usepackage{color}
\usepackage{comment}
\usepackage{xcolor}
\newcommand{\orcid}[1]{\href{https://orcid.org/#1}{\includegraphics[width=8pt]{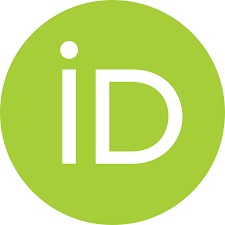}}}
\usepackage{nameref}
\usepackage{tikz}

\begin{document}
%opening
\title{Shadow of Bonanno–Reuter Black Hole in Plasma Medium: Insights from\\ EHT Sgr A* Observations}

\author{Shubham Kala\,\orcid{0000-0003-2379-0204}}
\email{shubhamkala871@gmail.com}
\affiliation{The Institute of Mathematical Sciences, C.I.T Campus, Taramani-600113, Chennai, Tamil Nadu, India }

\begin{abstract}
We investigate the properties of black hole shadows in the renormalization group (RG) improved Bonanno–Reuter spacetime, incorporating quantum gravitational corrections via the scale-dependent parameter $(\tilde{\omega})$ in a plasma medium. Light propagation in a non-uniform, pressureless plasma with a radial density profile is analyzed through modified equations of motion. The black hole shadow angular radius is computed, and its dependence on $\tilde{\omega}$ and the plasma index is analyzed. The analysis of specific limiting cases indicates systematic deviations of the black hole shadow relative to the classical Schwarzschild limit. Using Event Horizon Telescope (EHT) observations of Sgr~A*, we place constraints on $\tilde{\omega}$. Furthermore, within the considered parameter range, plasma and quantum-gravity effects exhibit an observational degeneracy, which future high-resolution measurements with the next-generation EHT are expected to break, thereby providing tighter constraints on the model parameters.\\

\end{abstract}

\maketitle

\section{Introduction}\label{sec1}

\noindent Black holes (BHs), which are solutions of general relativity (GR), are among the most fascinating objects in the universe \cite{Einstein:1916vd,Schwarzschild:1916uq}. The study of BH shadow has become a crucial method for exploring the strong gravitational field near compact objects, offering insights into spacetime geometry and fundamental physics. When light from an accretion disk or background source approaches a BH, gravitational lensing causes it to bend around the compact object~\cite{Batic:2014loa,Kala:2020viz,Kala:2021ppi,Kala:2025iri}, creating a dark region on the observer's sky known as the shadow \cite{chandrasekhar1998mathematical,Cunha:2019hzj}. The BH shadow represents the projected silhouette of the photon region as seen by a distant observer, shaped by the underlying spacetime curvature near the event horizon~\cite{Kala:2020prt}. Theoretical work by Synge \cite{Synge:1966okc} and Bardeen \cite{Bardeen:1973tla} laid the foundation for understanding shadow formation in Schwarzschild and Kerr geometries. The influence of charge and spin on the BH shadows was explored by Takahashi \cite{Takahashi:2005hy}, demonstrating notable deviations from the standard Schwarzschild BH (SBH) shadow. More recent studies by Falcke et al.~\cite{Falcke:1999pj}, Johannsen et al.~\cite{Johannsen:2010ru}, and Konoplya et al.~\cite{konoplya2019shadow,Konoplya:2021slg} have extended the theoretical framework of BH shadows to include realistic astrophysical environments and to test deviations from GR in alternative theories of gravity.

The breakthrough observational confirmation came with the EHT collaboration, which imaged the shadow of the supermassive BH M87* (Akiyama et al., 2019), providing the first direct evidence of horizon-scale structure \cite{EventHorizonTelescope:2019dse}. This achievement has sparked widespread interest in shadow-based tests of gravity, with numerous theoretical studies examining how modifications to GR, such as those arising from scalar-tensor theories, nonlinear electrodynamics, or extra-dimensional models, affect the size and shape of BH shadows \cite{Cunha:2018cof,Vagnozzi:2022moj}. In particular, the shadow's sensitivity to spacetime geometry makes it a valuable probe for constraining BH parameters and for distinguishing between different gravitational theories in future high-resolution observations.

In realistic astrophysical environments, BHs are rarely isolated; they are often surrounded by hot, ionized matter in the form of plasma. This plasma may arise from accretion flows, stellar winds, or interstellar media, and its distribution can significantly affect the propagation of light in the vicinity of the BH~\cite{breuer1980propagation}. Since plasma introduces a frequency-dependent refractive index, it modifies null geodesics and breaks the equivalence between geometrical and optical paths \cite{kulsrud1992dynamics}. Several studies, including those by Perlick et al.~\cite{Perlick:2015vta,Perlick:2023znh,Perlick:2025xvg} and Rogers \cite{rogers2015frequency}, have extended the theory of gravitational lensing to include dispersive effects of plasma, showing that the presence of plasma alters the apparent position and brightness of lensed images, as well as the shape of the BH shadow~\cite{Schee:2017hof,Ovgun:2020gjz,Ditta:2023rhr,Pahlavon:2024caj,Feleppa:2024vdk,Roy:2025qmx,Kobialko:2025sls}.

Building on this, various authors have investigated the influence of plasma environments, both homogeneous and inhomogeneous, on BH shadows in the vicinity of different BH solutions arising from GR as well as alternative theories of gravity. Studies by Bisnovatyi-Kogan et al. \cite{Bisnovatyi-Kogan:2017kii}, Attamurotov et al. \cite{Atamurotov:2015nra,atamurotov2023quantum,Turakhonov:2025ojy,Pahlavon:2024caj}, Övgün et al. \cite{Cimdiker:2021cpz,ali2025exploring,Ali:2024cti} and Perlick  Tsupko \cite{Perlick:2015vta}, Chowdhuri et al. \cite{Chowdhuri:2020ipb} and Kala et al. \cite{Kala:2022uog,Kala:2024fvg,Kala:2025fld} have shown that such plasma profiles can either enlarge or shrink the BH shadow depending on the photon frequency and plasma parameters. These findings underscore the importance of incorporating plasma effects when modeling the optical appearance of BHs, especially in the context of interpreting high-resolution observations from instruments like the EHT.

RG improved BH solutions incorporate quantum gravity (QG) corrections by allowing gravitational couplings to run with scale \cite{Bonanno:2000ep}. In asymptotically safe gravity (ASG), these corrections modify classical metrics at high energies while recovering GR in the infrared \cite{Platania:2023srt,Godani:2023jhq}. Quantum-corrected Kerr geometries were studied by Reuter and Tuiran \cite{Reuter:2010xb}, and thermodynamic implications by Becker and Reuter \cite{Becker:2012js}. RG improvements in scenarios with large extra dimensions were explored by Burschil and Koch \cite{Burschil:2009va}. Recent works focus on observational signatures: shadows under accretion \cite{Chen:2022dgn}, rotating cases \cite{Sanchez:2024sdm}, plasma environments \cite{Gohain:2025rip}, and quasinormal modes \cite{Konoplya:2023aph}, highlighting the phenomenological significance of RG improved models. Investigating the shadow of RG improved BHs in the presence of an astrophysical plasma medium is important for understanding how QG effects manifest in realistic observational settings. Plasma influences the propagation of light, and combined with deviations from GR, may lead to distinctive shadow features detectable by current or future telescopes.

The paper is organized as follows: In Section \ref{sec2}, we introduce the RG group improved BH and its geometry. In Section \ref{sec3}, we formulate the governing equations for the propagation of the light in plasma medium. Section \ref{sec4} focuses on the BH shadow and the constraints on its parameters based on EHT observations of Sgr A*.  Finally, we present our conclusions in Section \ref{sec5}.

\section{Bonanno--Reuter Black Hole} \label{sec2}
The Bonanno--Reuter BH (BRBH) emerges from the RG improvement of the classical Schwarzschild solution within the framework of ASG. In this approach, quantum gravitational effects are incorporated through a scale-dependent Newton's constant \( G(k) \), where \( k \) is an energy scale identified with the inverse of the radial coordinate \( r \). The effective Newton’s constant is modeled as \cite{Bonanno:2000ep}
\begin{equation}
G(r) = \frac{G_0\, r^3}{r^3 + \tilde{\omega} G_0 \left(r + \gamma G_0 M\right)},
\label{eq:effectiveG}
\end{equation}
where $G_0$ denotes the infrared (low-energy) Newton’s constant. 
The running coupling \( G(r) \), given in Eq.~\eqref{eq:effectiveG}, when substituted into the Schwarzschild lapse function \( f(r) = 1 - 2G(r)M/r \), leads to the RG-improved metric:
\begin{equation}
ds^2 = -f(r)\, dt^2 + \frac{dr^2}{f(r)} + r^2 d\Omega^2,
\end{equation}
with the lapse function explicitly given by \cite{Bolokhov:2025lnt},
\begin{equation}
f(r) = 1 - \frac{2 G_0 M r^2}{r^3 + \tilde{\omega} G_0 (r + \gamma G_0 M)}.
\label{eq:lapsefunction}
\end{equation}
Here, the parameter $\tilde{\omega}$ is treated as a free parameter and represents the first non-trivial QG correction appearing in the large-distance expansion of the RG improved coupling. The constant $\gamma$ is an interpolation parameter associated with the specific choice of cutoff identification in the RG improvement scheme. As noted in the original work \cite{Bonanno:2000ep}, setting $\gamma=0$ leads to essentially the same qualitative properties of the improved BH spacetime. Therefore, in the present work, we fix $\gamma$ to a positive constant and do not vary it in the shadow analysis, since it does not qualitatively affect the shadow characteristics.
This quantum-corrected BH geometry asymptotically approaches the Schwarzschild solution at large radial distances, with subleading quantum corrections of order $\mathcal{O}(1/r^{3})$. In the near-core region, the spacetime smoothly transitions to an effective de-Sitter geometry, characterized by an effective cosmological constant $\Lambda_{\mathrm{eff}} \sim 1/r_{0}^{2}$, thereby resolving the central curvature singularity. Depending on the values of the BH mass $M$ and the QG parameter $\tilde{\omega}$, the spacetime can exhibit two horizons (inner and outer), a single degenerate (extremal) horizon, or no horizon at all. For masses below a critical value $M_{\mathrm{cr}}$, no event horizon forms, leading to a horizonless, soliton-like remnant. Such features are well-known consequences of QG corrections and may give rise to potential observational signatures, including modifications of BH shadows and possible gravitational-wave echoes.
%%%%%%%%%%%%%%%%%%%%%%%%%%%%%%%%%%%%%%%%%%%%%%%%%%%%%%%%%%%%%%%%%%%%%%%%%%%%%%%%%
\begin{figure*} [tbhp]
	\centerline{
		\includegraphics[width=90mm,height=80mm]{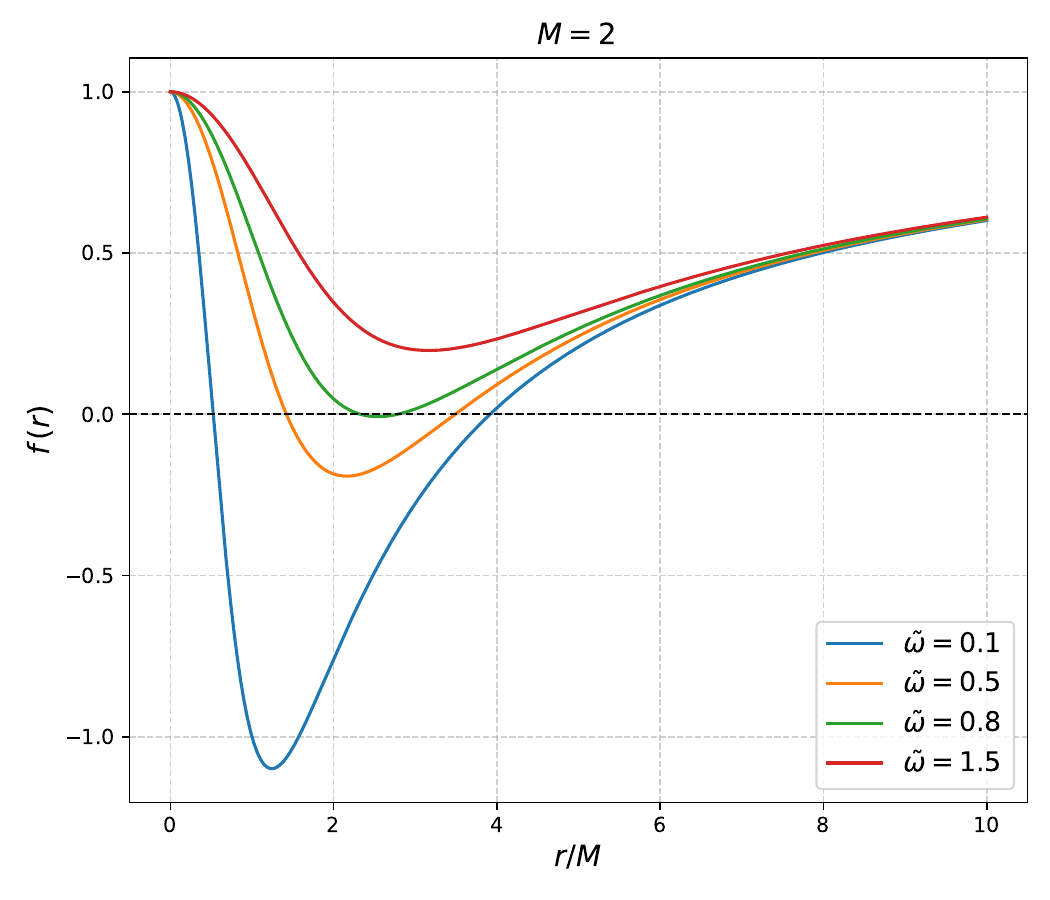}
        \includegraphics[width=90mm,height=80mm]{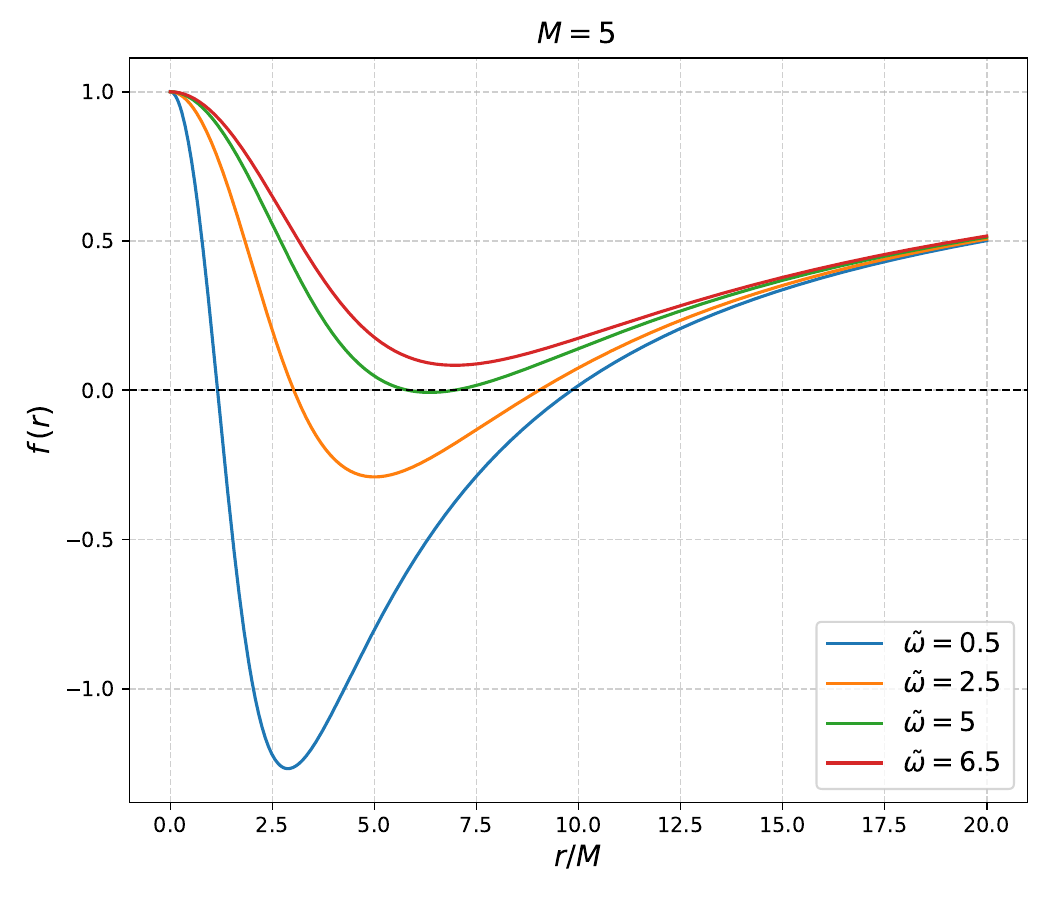}}
	\caption{Variation of the metric function with radial distance for different values of $\gamma$, considering (a) $M=2$ and (b) $M=3$. We set $G_{0}=1$ and \( \gamma=4.5 \).}
\label{fig:horizon}
\end{figure*}
%%%%%%%%%%%%%%%%%%%%%%%%%%%%%%%%%%%%%%%%%%%%%%%%%%%%
In Fig.~\ref{fig:horizon}, we plot the metric function $f(r)$ as a function of the radial coordinate $r$ for representative values of the mass parameter $M$ and the QG correction parameter $\tilde{\omega}$. The figure illustrates how variations in $\tilde{\omega}$ modify the causal structure of the BRBH spacetime. Specifically, depending on the value of $\tilde{\omega}$, the metric function admits different causal configurations, including spacetimes with two distinct horizons (inner and outer), a single degenerate (extremal) horizon, or no horizon at all. The sensitivity of the causal structure of the BRBH to the underlying parameters is consistent with earlier analyses reported in the literature \cite{Carballo-Rubio:2025fnc}. For each fixed value of $M$, increasing $\tilde{\omega}$ drives a smooth transition between these causal regimes, highlighting the role of QG corrections in deforming the classical SBH horizon structure. In particular, for $M = 2$, the transition from a two-horizon configuration to a horizonless spacetime occurs at approximately $\tilde{\omega} \simeq 0.8$, while for $M = 5$ this transition shifts to $\tilde{\omega} \simeq 5$. These values mark the boundary between BH and horizonless configurations for the corresponding masses.

\section{Photon Motion of Bonanno--Reuter BH in Plasma Medium} \label{sec3}
The Lagrangian describing a massless test particle (photon) moving in a general static, spherically symmetric spacetime and confined to the equatorial plane ($\theta = \pi/2$) is given by
\begin{equation}
    \mathcal{L} = \frac{1}{2} g_{\mu\nu} \dot{x}^{\mu} \dot{x}^{\nu}
    = \frac{1}{2} \left( g_{tt} \dot{t}^2 + g_{rr} \dot{r}^2 + g_{\phi\phi} \dot{\phi}^2 \right).
\end{equation}
In vacuum, this Lagrangian leads to null geodesic motion for photons. However, in the presence of a plasma medium, photon propagation is no longer governed by null geodesics of the spacetime metric alone. Instead, it is described by an effective Hamiltonian derived from the dispersion relation in a refractive medium. The corresponding Hamiltonian governing photon propagation in plasma is given by~\cite{Synge:1966okc},
\begin{equation}
    H = \frac{1}{2} \left( g^{\mu\nu} p_\mu p_\nu + \omega_p^2 \right),
\end{equation}
where $\omega_p$ denotes the plasma frequency, which characterizes the response of the surrounding ionized medium and depends on the local electron number density. For a cold, non-magnetized plasma, it is defined as
\begin{equation}
    \omega_p^2(r) = \frac{4\pi e^2}{m_e} N(r),
\end{equation}
with $e$ and $m_e$ being the electron charge and mass, respectively, and $N(r)$ the electron number density. In the homogeneous case, $\omega_p$ can be treated as a constant. However, in the present work, we consider a non-homogeneous plasma environment and model the plasma frequency using a radial power-law profile~\cite{rogers2015frequency},
\begin{equation}
    \omega_p^2(r) = \frac{h}{r^\beta},
\end{equation}
where $h$ and $\beta$ are constants with $\beta \geq 0$. The special case $\beta = 0$ corresponds to a homogeneous plasma. Throughout this work, we consider $\beta = 1$.
The presence of plasma introduces a frequency-dependent refractive index, given by
\begin{equation}
    n^2 = 1 - \left( \frac{\omega_p}{\omega} \right)^2,
\end{equation}
where $\omega$ denotes the photon frequency measured by a static observer at radius $r$. Owing to gravitational redshift, this frequency is related to the conserved frequency $\omega_0$ measured at spatial infinity as
\begin{equation}
    \omega = \frac{\omega_0}{\sqrt{-g_{00}}}.
\end{equation}
Thus, using $p_0 = E_0 = \omega_0$, the Hamiltonian can be rewritten as
\begin{equation}
    H = \frac{1}{2} \left( g^{\mu\nu} p_\mu p_\nu + (n^2 - 1)\, g^{00} p_0^2 \right).
\end{equation}
Applying Hamilton’s equations, $\dot{p}_\mu = -\partial H / \partial x^\mu$, and noting that the plasma frequency depends only on the radial coordinate, $\omega_p = \omega_p(r)$, we identify the conserved quantities as
\[
p_0 = -E, \qquad p_\phi = L,
\]
where $E$ and $L$ denote the conserved energy and angular momentum of the photon, respectively.
The canonical equations lead to the equations of motion in the equatorial plane, given by
\begin{align}
    \frac{dt}{d\lambda} &= \frac{n^2 E}{f(r)}, \\
    \frac{d\phi}{d\lambda} &= \frac{L}{r^2}, \\
    \left( \frac{dr}{d\lambda} \right)^2 &= n^2 E^2 - \frac{L^2}{r^2} f(r),
\end{align}
where $\lambda$ is an affine parameter along the photon trajectory, and $E$ and $L$ denote the conserved energy and angular momentum of the photon, respectively. Combining the above equations yields the trajectory equation
\begin{equation}
    \left( \frac{dr}{d\phi} \right)^2
    = r^4 \left[ \frac{n^2(r)\, E^2}{L^2} - \frac{f(r)}{r^2} \right].
\end{equation}
%%%%%%%%%%%%%%%%%%%%%%%%%%%%%%%%%%%%%%%%%%%%%%%%%%%%%%%%%%%%%%%%%%%%%%%%%%%%%%%%%
\begin{figure*} [tbhp]
	\centerline{
		\includegraphics[width=90mm,height=90mm]{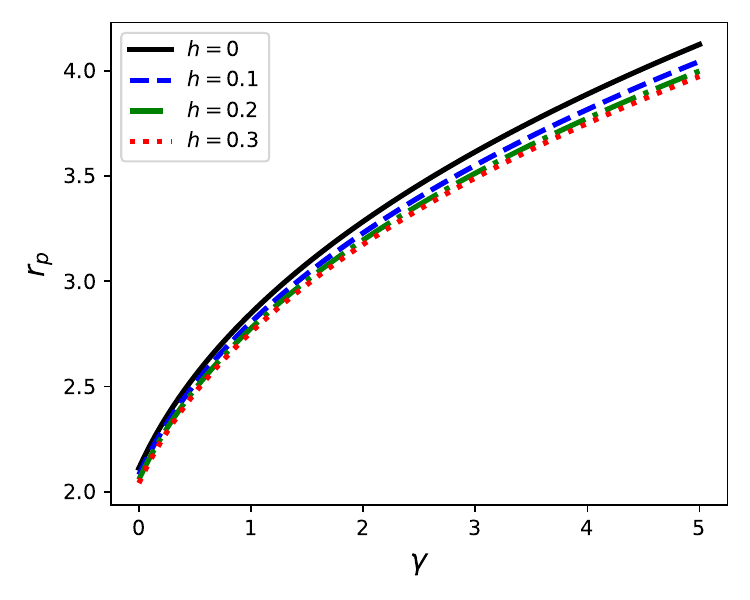}}
	\caption{Variation of radius of photon sphere as a function of $\gamma$ for different values of $h$.  Parameters are chosen as $M=5$, $G_{0}=1$, $k_{p}=0.5$, and \( \tilde{\omega} = 118/(15\pi) \).}
\label{Fig2B}
\end{figure*}
%%%%%%%%%%%%%%%%%%%%%%%%%%%%%%%%%%%%%%%%%%%%%%%%%%%%%%%%%%%%%%%%%%%%%%%%%%%%%%%%%
%%%%%%%%%%%%%%%%%%%%
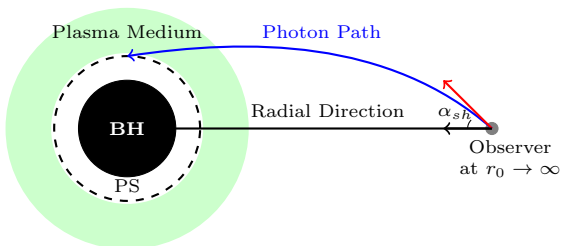
\begin{figure} [tbhp]
\centering
{\scriptsize
\begin{tikzpicture}[scale=0.8]

% Plasma medium (shaded annular region)
\fill[green!20] (0,0) circle (2.0cm); % Outer plasma boundary
\fill[white] (0,0) circle (1.25cm); % Inner cut-out inside plasma

% BHand photon sphere
\filldraw[black] (0,0) circle (0.8cm); % Black hole
\draw[dashed, thick] (0,0) circle (1.2cm); % Photon sphere

% Labels
\node[text=white] at (0,0) {\textbf{BH}}; % Inside black hole
\node at (0,-0.95) {PS};
\node at (0,1.6) {Plasma Medium}; % Moved upward

% Observer
\filldraw[gray] (6,0) circle (0.1cm);
\node[align=center] at (6.3,-0.5) {Observer \\ at $r_0 \to \infty$};

% Extended radial line from observer to near BH
\draw[->, thick] (6,0) -- (0.6,0);
\node at (3.3,0.3) {Radial Direction};

% Photon path (bent light)
\draw[blue, thick, ->] (6,0) .. controls (4.3,1.5) and (2,1.5) .. (0,1.2);
\node[blue] at (3.2,1.6) {Photon Path};

% Shadow angle alpha_sh
\draw[->, thick, red] (6,0) -- (5.2,0.8); % bent light
\draw[->, thick, black] (6,0) -- (5.2,0); % radial direction

% Angle alpha_sh arc
\draw (5.6,0) arc[start angle=180, end angle=135, radius=0.2cm];
\node at (5.4,0.25) {$\alpha_{sh}$};

\end{tikzpicture}
}
\caption{Schematic representation of photon deflection near a BH surrounded by a plasma medium. $\alpha_{sh}$ is the angular radius of the shadow~\cite{Bisnovatyi-Kogan:2017kii,Roy:2025hdw}.}
\label{fig:photon-deflection-plasma}
\end{figure}
%%%%%%%%%%%%%%%%%%%%%%%%%%%%%%%%%%%%%%%%%%%%%%%%%%%%%%%%%%%%%%%%%%%%%%%%%%%%%%%%%
\begin{figure*} [tbhp]
	\centerline{
		\includegraphics[width=90mm,height=90mm]{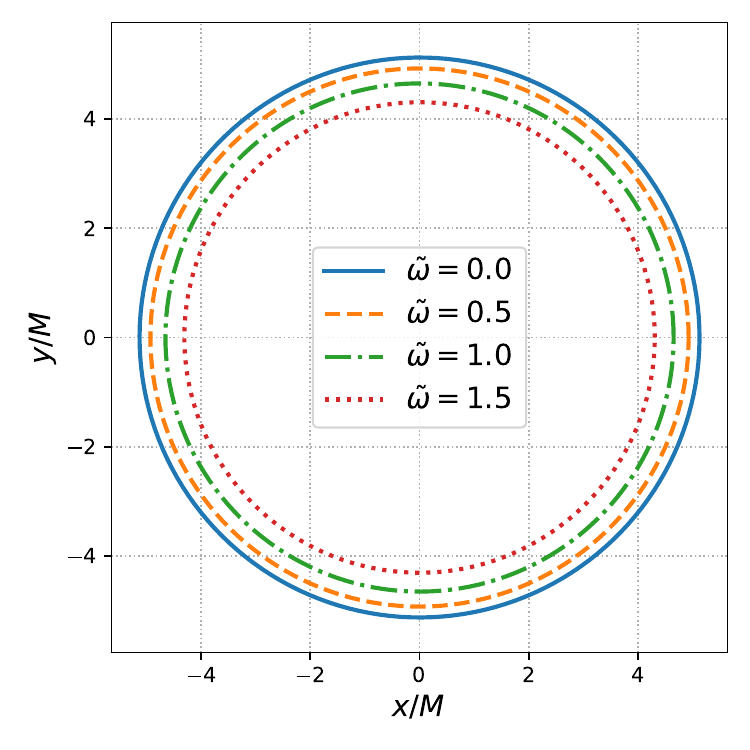}
        \includegraphics[width=90mm,height=90mm]{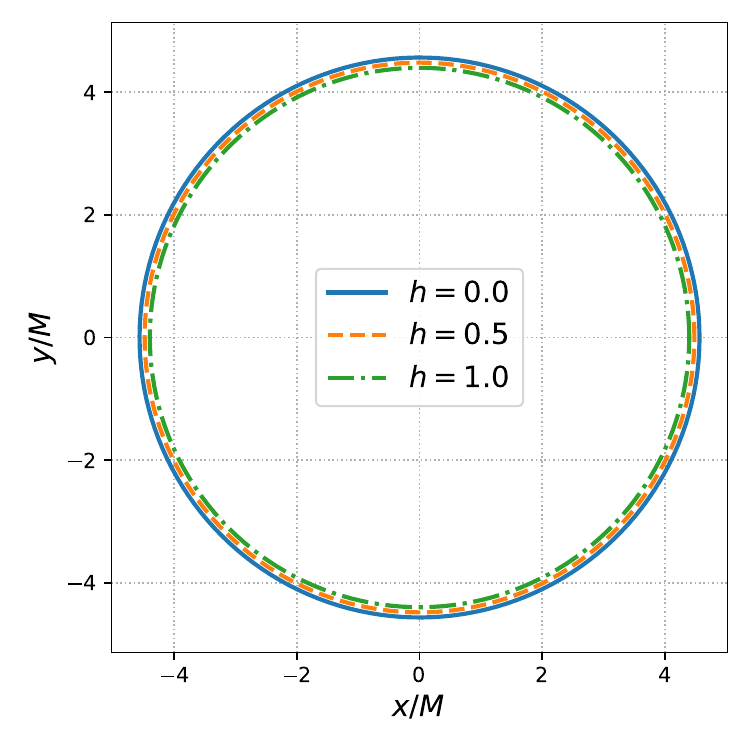}}
	\caption{The angular shadow radius of the BRBH:
(a) for different values of the QG correction parameter $\tilde{\omega}$, with the plasma index fixed at $h = 1$ and $\gamma = 4.5$; 
(b) for different values of the plasma index parameter $h$, with $\tilde{\omega} = 0.5$ and $\gamma = 4.5$ fixed. 
Here, $x/M$ and $y/M$ denote the angular celestial coordinates on the observer’s sky, normalized by the BH mass $M$ (or equivalently by the shadow radius $R_{\rm sh}/M$).}
\label{Fig1}
\end{figure*}
%%%%%%%%%%%%%%%%%%%%%%%%%%%%%%%%%%%%%%%%%%%%%%%%%%%%
%%%%%%%%%%%%%%%%%%%%%%%%%%%%%%%%%%%%%%%%%%%%%%%%%%%%%%%%%%%%%%%%%%%%%%%%%%%%%%%%%
\begin{figure*} [tbhp]
	\centerline{
		\includegraphics[width=120mm,height=95mm]{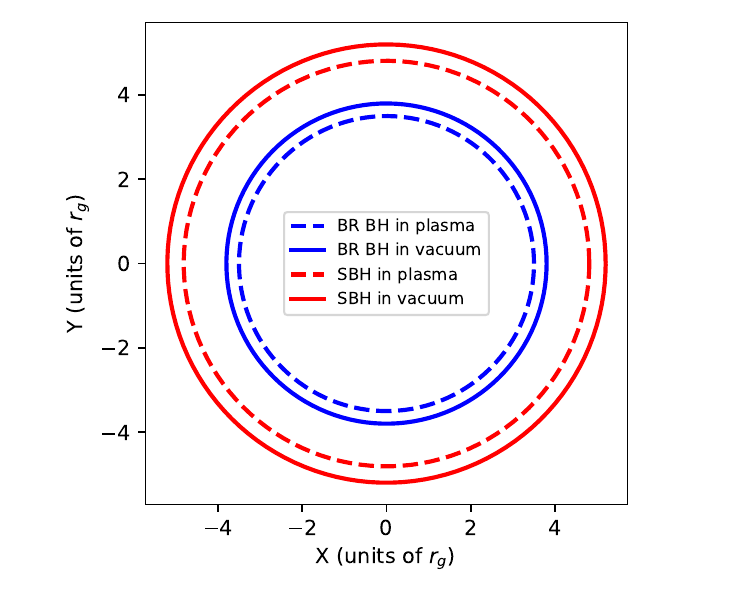}}
	\caption{Comparison of the shadows of the BRBH and SBH, each computed in both vacuum and in the presence of a plasma medium. The parameters are chosen as $M=1$ for the SBH, $M=2$ for the BRBH, $G_0=1$, $\tilde{\omega}=0.5$, plasma density falloff index $h=3$ where applicable, and the parameter $\gamma=4.5$ fixed for the BRBH.  
Here, $X$ and $Y$ denote the angular celestial coordinates on the observer’s sky, normalized by the BH mass $M$ (or equivalently by the shadow radius $R_{\rm sh}/M$).}
\label{Fig2}
\end{figure*}
%%%%%%%%%%%%%%%%%%%%%%%%%%%%%%%%%%%%%%%%%%%%%%%%%%%%
%%%%%%%%%%%%%%%%%%%
The radius of the photon sphere, $r_p$, in the plasma medium is determined by the existence of unstable circular photon orbits. These satisfy the conditions
\[
\dot r = 0
\quad \text{and} \quad
\frac{d}{dr}\left(\dot r^{\,2}\right) = 0,
\]
evaluated at $r = r_p$. Here, the dot denotes differentiation with respect to the affine parameter. Applying these conditions to the radial equation leads to the photon sphere equation
\begin{equation} \label{EqPS1}
\left.
\frac{d}{dr}\left(\frac{n^2(r)\, r^2}{f(r)}\right)
\right|_{r = r_p} = 0 .
\end{equation}
Equivalently, this condition can be written in the explicit form
\begin{equation} \label{EqPS2}
\left.
\left[
2 n(r) n'(r)\, r\, f(r)
+ 2 n^2(r) f(r)
- n^2(r)\, r\, f'(r)
\right]
\right|_{r = r_p} = 0 .
\end{equation}
The radius of the photon sphere, obtained by numerically solving Eq.~\ref{EqPS2}, is plotted in Fig.~\ref{Fig2B} as a function of the parameter $\gamma$ for different values of the plasma exponent $h$. It is clearly observed that the radius of the photon sphere increases with increasing $\gamma$. Furthermore, in the absence of the plasma medium (that is, for negligible plasma density), the radius of the photon sphere attains its maximum value and decreases as the plasma strength parameter $h$ increases.
%%%%%%%%%%%%%%%%%%%%%%%%%%%%%%%%%%%%%%%%%%%%%%%%%%%%%%%%%%%%%%%%%%%%%%%%%%%%%%%%%%%%
\section{Shadow of the Black Hole} \label{sec4}
Astrophysical BHs are often surrounded by ionized matter, forming accretion disks and diffuse plasma clouds~\cite{Uzdensky:2014rza}. The presence of such plasma environments modifies the propagation of photons through a frequency-dependent refractive index, leading to observable changes in the apparent shadow of the BH. To account for these effects, we consider the propagation of light rays confined to the equatorial plane of the BRBH in the presence of a cold, non-magnetized and inhomogeneous plasma. The schematic figure representing this setup is shown in Fig.~\ref{fig:photon-deflection-plasma}. Following the standard geometric approach developed by Perlick et al.~\cite{Perlick:2015vta}, the angular radius of the shadow as seen by a static observer located at radial coordinate $r_0$ is given by
\begin{equation} \label{BHSEq}
R_{sh} = \frac{r_p}{\sqrt{f(r_p)}}.
\end{equation}

Equation~\eqref{BHSEq} allows us to systematically examine how the shadow of the BRBH varies with the QG correction parameter~$\tilde{\omega}$ and the surrounding plasma environment. The corresponding results are shown in Fig.~\ref{Fig1}. In subfigure~(a) of Fig.~\ref{Fig1}, we fix the plasma parameters and vary~$\tilde{\omega}$. The results show that the shadow radius decreases monotonically with increasing~$\tilde{\omega}$. This behavior can be physically interpreted as a manifestation of the running of the gravitational coupling constant in the ASG framework: higher values of~$\tilde{\omega}$ enhance the effective QG corrections near the horizon, leading to a reduction in the effective gravitational radius and consequently a smaller photon sphere. In other words, the QG corrections act to weaken the spacetime curvature in the near-horizon region, resulting in a reduced apparent shadow size as perceived by a distant observer.
Similarly, in subfigure~(b) of Fig.~\ref{Fig1}, we fix~$\tilde{\omega}$ and vary the plasma strength. Here, we find that the shadow radius systematically decreases with stronger plasma effects. This behavior arises because a denser plasma increases the local refractive index, which modifies the null geodesics followed by light rays and effectively shrinks the observed angular size of the shadow.\\
%%%%%%%%%%%%%%%%%%%%%%%%%%%%%%%%%%%%%%%%%%%%%%%%%%%%%%%%%%%%%%%%%%%%%%%%%%%%%%%%%
\begin{figure*} [tbhp]
	\centerline{
		\includegraphics[width=90mm,height=90mm]{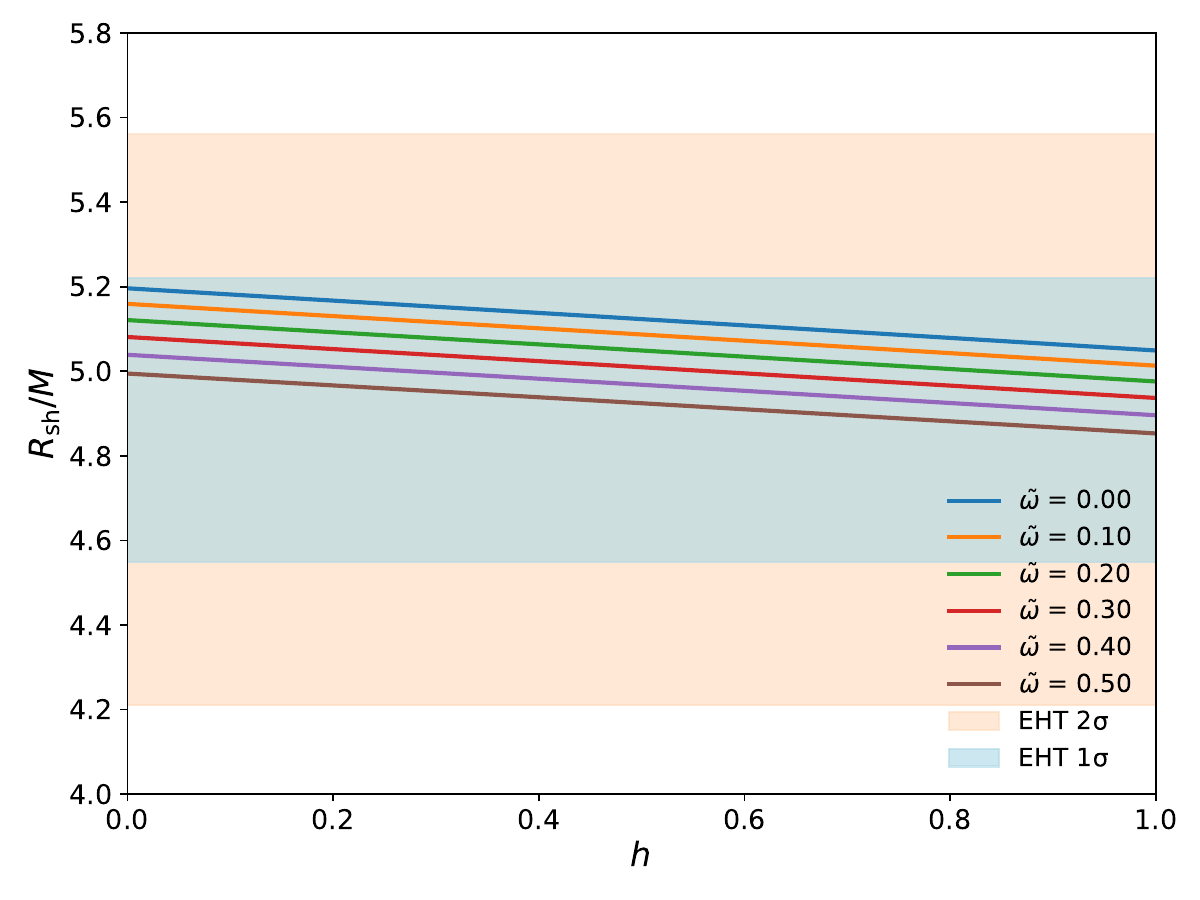}
        \includegraphics[width=90mm,height=90mm]{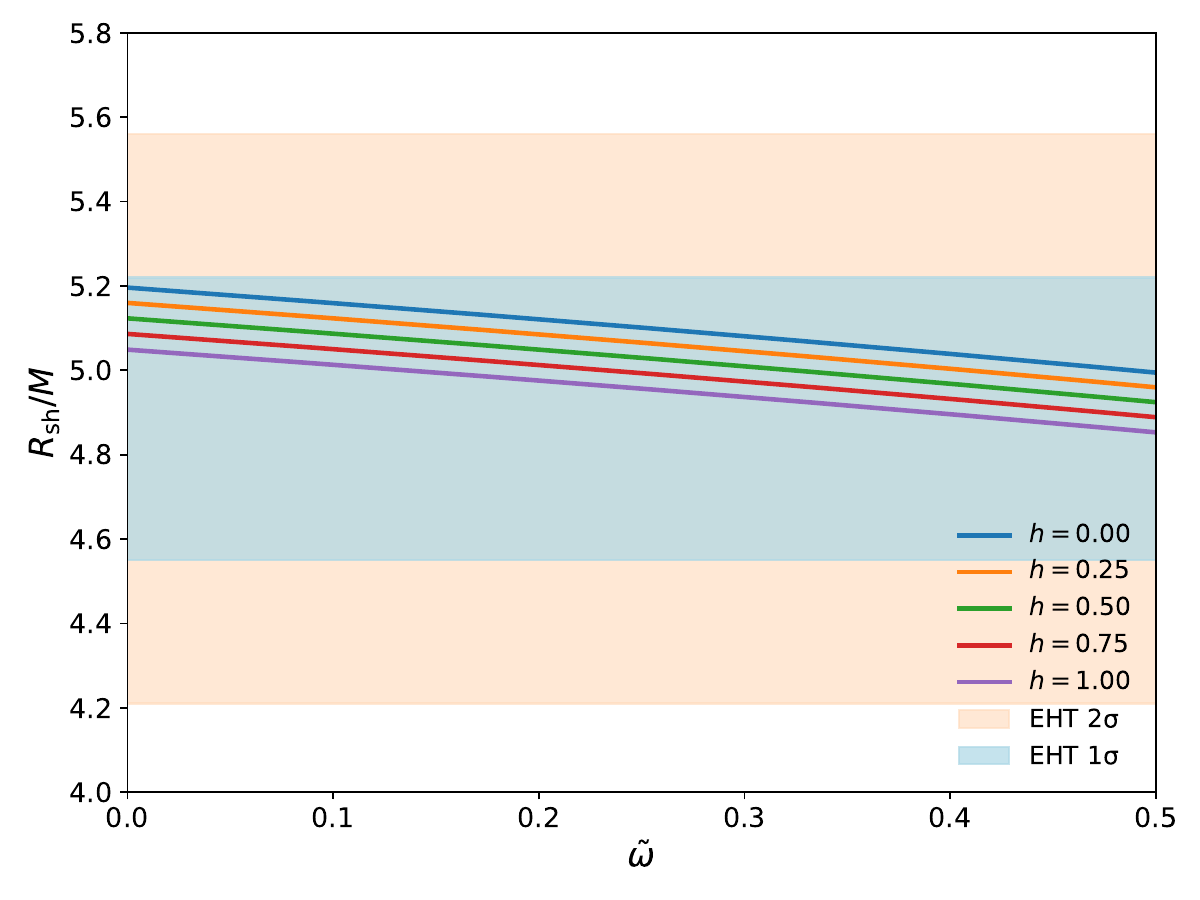}}
	\caption{Shadow radius of the BRBH compared with the EHT observational constraints for Sgr~A*, including the $1\sigma$ and $2\sigma$ confidence regions.}
\label{Fig4}
\end{figure*}
The angular shadow size of the BRBH in plasma reduces to several well-known limiting cases within specific parameter ranges. For instance, setting $h = 0$ reproduces the BRBH in vacuum, setting $\tilde{\omega} = 0$ reproduces the SBH in plasma, and setting both $h = 0$ and $\tilde{\omega} = 0$ reproduces the classical SBH in vacuum. In Fig.~\ref{Fig2}, we present a qualitative comparison of the four scenarios based on these theoretical results. In the parameter range considered, the shadow sizes decrease successively from the SBH in vacuum to the SBH in plasma, to the BRBH in vacuum, and finally to the BRBH in plasma.
This systematic deviation arises from two physical effects. The presence of plasma reduces the shadow size by decreasing the effective speed of light, while quantum corrections in the BRBH reduce the radius of the photon sphere due to the scale dependence of the Newton constant in the RG-improved geometry. Within the considered parameter range, the combined influence of plasma and quantum corrections results in the smallest shadow for the BRBH in plasma. We note that although this behavior is well defined for the chosen parameter values, a broader exploration of plasma densities and quantum correction parameters may lead to partial degeneracies in the observed shadow sizes.\\
We extend our analysis by constraining the shadow results using observational data from the EHT for Sgr A*. The angular radius of the observed shadow lies within the $1\sigma$ band of $4.55$–$5.22$ and the $2\sigma$ band of $4.25$–$5.56$ in units of $GM/c^2$ \cite{EventHorizonTelescope:2022wkp,Kala:2025xnb}. We compare our theoretical predictions for the photon sphere radius, considering different plasma profiles and values of the QG correction parameter $\tilde{\omega}$, with these observational limits. The agreement within the $1\sigma$ and $2\sigma$ confidence intervals serves as a consistency check and helps to assess the physical viability of the proposed model.\\

Fig.~\ref{Fig4} illustrates the normalized shadow radius $R_{\rm sh}/M$ of the BRBH in a plasma environment with QG corrections, compared with the EHT $1\sigma$ and $2\sigma$ observational bounds for Sgr~A*. All theoretical curves lie within the $1\sigma$ confidence region, which is itself contained within the $2\sigma$ region, indicating that for the considered ranges of the plasma index ($0 \le h \le 1.0$) and the QG parameter ($0 \le \tilde{\omega} \lesssim 0.5$), neither plasma refraction nor quantum corrections lead to shadow sizes inconsistent with current observations. The shadow radius decreases monotonically with increasing plasma index or QG strength, reflecting enhanced light refraction in a denser plasma and modifications of photon trajectories near the BH due to quantum effects. The potential observational degeneracy between plasma and quantum effects is a notable feature, whereby an increase in plasma density can compensate for a smaller quantum correction, and vice versa, leading to similar shadow sizes in observations. However, within the specific limits considered in Fig.~\ref{Fig2}, a clear and systematic deviation in the shadow sizes is obtained. This comparison is based on a theoretical limiting analysis designed to isolate the deviation from the classical SBH case due to individual plasma and QG parameter. While current EHT observations cannot independently distinguish between plasma-induced and QG-induced modifications, future high-resolution measurements with the next-generation Event Horizon Telescope (ngEHT) are expected to resolve this observational degeneracy and place tighter constraints on the BRBH parameters.
\begin{table} [tbhp]
\centering
\caption{Upper bounds on $\tilde{\omega}$ from EHT observations for different plasma indices $h$.}
\label{tab:omega_constraints}
\begin{tabular}{c c c}
\hline
$h$ & Upper 1$\sigma$ & Upper 2$\sigma$ \\
\hline
0.00 & 1.1791 & 1.1908 \\
0.10 & 1.1734 & 1.5703 \\
0.20 & 1.1607 & 1.5149 \\
0.30 & 1.1639 & 1.5144 \\
0.40 & 1.1375 & 1.5703 \\
0.50 & 1.1247 & 1.5325 \\
%0.60 & 1.1109 & 1.5608 \\
%0.70 & 1.0963 & 1.5401 \\
%0.80 & 1.0809 & 3.9061 \\
%0.90 & 1.0645 & 1.5487 \\
%1.00 & 1.0473 & 3.6506 \\
\hline
\end{tabular}
\end{table}
The numerical analysis provides upper bounds on the QG parameter $\tilde{\omega}$ for different plasma indices $h$ based on the EHT observational constraints for Sgr~A*. The results are summarized in Table~\ref{tab:omega_constraints}. For $h = 0.0$, the shadow radius allows $\tilde{\omega} \lesssim 1.179$ at the 1$\sigma$ confidence level and $\tilde{\omega} \lesssim 1.191$ at 2$\sigma$. As the plasma index increases, the allowed $\tilde{\omega}$ gradually decreases at 1$\sigma$, reaching $\tilde{\omega} \lesssim 1.047$ for $h = 1.0$, while the 2$\sigma$ bounds vary more irregularly due to wider observational uncertainties, with $\tilde{\omega} \lesssim 3.651$ at $h = 1.0$. This indicates that for moderate plasma densities ($h \lesssim 0.5$), the BRBH remains fully consistent with EHT observations, and even at higher plasma indices, the model predictions stay largely within the 2$\sigma$ range, highlighting the robustness of the constraints and the degeneracy between plasma effects and QG corrections.\\
We compare our results for a consistency check with the study by Kumar et al.~\cite{Kumar:2019ohr}, which analyzed the shadow of rotating BHs in ASG. In the limit $\tilde{\omega} = 0$ and in the absence of plasma, our findings qualitatively reproduce the shadow radius and shape trends reported by Kumar et al., confirming consistency with their results. While they observed that the shadow radius decreases and intrinsic distortion increases with larger $\zeta$, we find that increasing the QG parameter $\tilde{\omega}$ similarly reduces the shadow radius. Our work further extends this analysis by including plasma effects, demonstrating that refractive modifications also shrink the apparent shadow size, thereby highlighting the combined impact of QG corrections and environmental effects. Overall, this comparison establishes that our framework is consistent with previous studies while providing additional insights for plasma environments.

%%%%%%%%%%%%%%%%%%%%%%%%%%%%%%%%%%%%%%%%%%%%%%%%%%
\section{Conclusions} \label{sec5}
\noindent In this paper, we investigated the shadow properties of a BRBH in the presence of a non-uniform, magnetized, pressureless cold plasma. The key results are as follows:
\begin{itemize}    
    \item The shadow radius of the BRBH decreases monotonically with increasing QG correction parameter $\tilde{\omega}$, as higher $\tilde{\omega}$ strengthens quantum corrections near the horizon, reducing the effective photon sphere. Similarly, the shadow radius decreases with increasing plasma index $h$, reflecting enhanced light refraction in a denser plasma. These effects highlight the combined role of spacetime geometry and medium properties in shaping the shadow.
    \item The angular shadow size of the BRBH in plasma reduces to the corresponding well-known limiting cases, and a qualitative comparison of these scenarios is presented. Within the considered parameter range, the combined effects of plasma and quantum corrections yield the smallest shadow for the BRBH in plasma, clearly demonstrating the deviation from the classical limit.
    \item The QG correction parameter $\tilde{\omega}$ is constrained using EHT observations of Sgr~A* for different plasma indices $h$. The results indicate that, for moderate plasma densities and quantum corrections, the predicted shadow radius remains within the $1\sigma$ confidence region. The upper bounds on $\tilde{\omega}$ decrease slightly as $h$ increases, while the $2\sigma$ limits remain more relaxed due to larger observational uncertainties (see Table~\ref{tab:omega_constraints}).
    \item The combined influence of plasma effects and QG corrections can lead to an observational degeneracy when constraining the model with EHT data, since an increase in plasma density may compensate for a smaller quantum correction, making it difficult to disentangle the individual effects of these parameters. Future high-resolution observations with the ngEHT are expected to break this degeneracy and place tighter constraints on both the QG parameter $\tilde{\omega}$ and the plasma properties.
    \item Finally, we compare our results with those of Kumar et al.~\cite{Kumar:2019ohr}, who analyzed shadows of rotating BHs in ASG. In the limit $\tilde{\omega} = 0$ and in the absence of plasma, our results qualitatively match their findings.
\end{itemize}

\noindent These results show that both quantum gravitational corrections and plasma effects influence the size of BH shadows. Within the studied parameter range, the shadow sizes follow a predictable pattern, but in a broader parameter space, degeneracies can occur, which may affect observational interpretation. This analysis highlights the need to account for both quantum gravity and astrophysical environments when interpreting horizon-scale BH observations.
%%%%%%%%%%%%%%%%%%%%%%%%%%%%%%%%%%%%%%%%%%%%%%%%%%%%
\section*{Acknowledgments}
\noindent We thank the referee for the insightful comments, which have significantly improved the presentation and consistency of the manuscript. The author acknowledges the Institute of Mathematical Sciences (IMSc), Chennai for providing excellent research facilities and a conducive environment that supported his work as an Institute Postdoctoral Fellow. The author also gratefully acknowledges the contribution of COST Action CA21136 – “Addressing Observational Tensions in Cosmology with Systematics and Fundamental Physics (CosmoVerse).”
%\section*{Conflict of Interest}
%The authors declare no conflict of interest.
%\section*{Data Availability Statement}
%No datasets were generated during the current study.
%%%%%%%%%%%%%%%%%%%%%%%%%%%%%%%%%%%%%%%%%%%%%%%%%%%%%%%%%%%%%%%%%%%%%%%%%%%%%%
\bibliography{mainBRBH}
\bibliographystyle{unsrt}
\end{document}